\documentclass[%
 aip,
 apl,
 amsmath,amssymb,
 reprint,%
noeprint]{revtex4-1}

\usepackage{graphicx}
\usepackage{dcolumn}
\usepackage{bm}
\usepackage{mathptmx}
\usepackage{etoolbox}
\usepackage{braket}
\usepackage{hyperref}
\hypersetup{
    colorlinks=true,
    linkcolor=blue,     
    urlcolor=blue,
    citecolor=blue,
}
\usepackage[capitalise]{cleveref}
\usepackage{multirow}
\usepackage{tabularx}
\usepackage[euler]{textgreek}
\usepackage{xcolor}
\usepackage[version=4]{mhchem}
\usepackage{float}
\usepackage[mathlines]{lineno}
\usepackage{comment}
\usepackage{mathtools}

\DeclarePairedDelimiter\abs{\lvert}{\rvert}
\DeclarePairedDelimiter\norm{\lVert}{\rVert}

\makeatletter
\def\@email#1#2{%
 \endgroup
 \patchcmd{\titleblock@produce}
  {\frontmatter@RRAPformat}
  {\frontmatter@RRAPformat{\produce@RRAP{*#1\href{mailto:#2}{#2}}}\frontmatter@RRAPformat}
  {}{}
}%
\makeatother
\begin{document}

\title[Widefield NV Magnetic Field Reconstruction]{Widefield NV Magnetic Field Reconstruction for Probing the Meissner Effect and Critical Current Density under Pressure}
\author{Kin On Ho}
 \email{kin.ho@ens-paris-saclay.fr}
 \altaffiliation[Present address: ]{Department of Physics, The University of Texas at Austin, Austin, TX, USA}
 \affiliation{ 
Université Paris-Saclay, CNRS, ENS Paris-Saclay, CentraleSupelec, LuMIn, F-91190 Gif-sur-Yvette, France
}%
\author{Cassandra Dailledouze}
 \affiliation{ 
Université Paris-Saclay, CNRS, ENS Paris-Saclay, CentraleSupelec, LuMIn, F-91190 Gif-sur-Yvette, France}%
\author{Martin Schmidt}
 \email{martin.schmidt@universite-paris-saclay.fr}
 \affiliation{ 
Université Paris-Saclay, CNRS, ENS Paris-Saclay, CentraleSupelec, LuMIn, F-91190 Gif-sur-Yvette, France}%
\author{Loïc Toraille}
 \affiliation{CEA DAM, DIF, F-91297 Arpajon, France}
 \affiliation{Université Paris-Saclay, CEA, Laboratoire Matière en Conditions Extrêmes, 91680 Bruyères-le-Châtel, France}
\author{Marie-Pierre Adam}
 \affiliation{ 
Université Paris-Saclay, CNRS, ENS Paris-Saclay, CentraleSupelec, LuMIn, F-91190 Gif-sur-Yvette, France}%
\author{Jean-François Roch}
 \email{jean-francois.roch@ens-paris-saclay.fr}
 \affiliation{ 
Université Paris-Saclay, CNRS, ENS Paris-Saclay, CentraleSupelec, LuMIn, F-91190 Gif-sur-Yvette, France}%

\date{\today}

\begin{abstract}
The spatial distribution of a magnetic field can be determined with micrometer resolution using widefield nitrogen vacancy (NV) center magnetic imaging. Nevertheless, reconstructing the magnetic field from the raw data can be challenging due to the degeneracy of the four possible NV axes and the tremendous amount of data. While a qualitative approach is sufficient for most analyses, a quantitative analysis offers deeper insight into the physical system. Here, we apply NV widefield magnetic imaging to a \ce{HgBa2Ca2Cu3O_{8+\delta}} (Hg-1223) superconducting microcrystal at a pressure of 4 GPa. We fit the results with solutions from the Hamiltonian describing the NV center ground state and take into account the relative intensities of the resonances to determine the local magnetic field magnitude and angle. Thus, we reconstruct the temperature-dependent expulsion of the magnetic field associated with the Meissner effect around the superconductor. By comparing the resulting parameters to Brandt's model, which describes the magnetic behavior of a type-II superconductor, we extract the critical current density $j_c$. Overall, this work showcases the first widefield quantitative reconstruction of the field screening under pressure and an optical method to study critical current density. Thus, it provides new insights into the application of NV magnetometry to superconductivity research at high pressures.
\end{abstract}

\maketitle

The Meissner effect, whereby an external magnetic field is expelled from the interior of a superconductor, is a defining feature of superconductivity. The spatial distribution of the magnetic field around a superconductor contains a large amount of information, particularly about the superconducting currents responsible for the Meissner effect. However, obtaining this spatial map is nontrivial, and the situation worsens under pressure. Most conventional techniques only detect a bulk response and are fragile at high pressures. This hinders a thorough analysis of complex materials, such as superhydrides and bulk nickelates~\cite{Mao2018Solids, Boeri2022The, Ho2024Quantum}.

Nitrogen-vacancy (NV) magnetometry is a promising tool for studying condensed matter physics~\cite{Rondin2014Magnetometry, Casola2018Probing, Acosta2019Color, Ho2021Diamond, Ho2021Recent, Xu2023Recent}. Over the past decade, significant efforts have been devoted to use NV centers as a magnetic probe for superconductivity at both ambient pressure~\cite{Bouchard2011Detection, Waxman2014Diamond, Thiel2016Quantative, Pelliccione2016Scanned, Nusran2018Spatially, Schlussel2018Wide, Rohner2018Real, Joshi2019Measuring, Xu2019Mapping, Lillie2020Laser, Joshi2020Quantum, McLaughlin2021Strong, Paone2021All, Scheidegger2022Scanning, Monge2023Spin, Nishimura2023Wide, Ho2025Studying, Liu2025Quantum, Jayaram2025Probing} and high pressures obtained in a diamond anvil cell (DAC)~\cite{Yip2019Measuring, Lesik2019Magnetic, Hsieh2019Imaging, Bhattacharyya2024Imaging, Dailledouze2025Imaging, Wen2025Imaging}. Interestingly, the vector information of a magnetic field is encoded in the optically detected magnetic resonance (ODMR) spectrum of an ensemble of NV centers under microwave (MW) excitation, resulting from the projection of this vector on the four crystallographic N-to-V axes~\cite{Chipaux2015Wide}. It is possible to perform a fast and robust qualitative analysis using a [100]-cut diamond to obtain the superconducting properties of \ce{HgBa2Ca2Cu3O_{8+\delta}} (Hg-1223) under pressure, as was done in Ref.~\cite{Dailledouze2025Imaging}. However, quantitatively mapping the Meissner effect is more challenging. In this study, we present a phenomenological model aimed at quantitatively reconstructing the magnetic field vector of the field screening under pressure. We then use this model to revise the experimental data and simulate ODMR spectra to determine the temperature dependence of the magnetic field texture around the superconductor. The spatially reconstructed magnetic fields allow us to use Brandt's model to determine the temperature-dependent critical current density $j_{c}$, which is the maximum current density that the Hg-1223 sample can carry before losing its superconductivity. Thus, we have developed an optical method to reveal the temperature-dependent critical current density $j_{c}$ under pressure, thereby mitigating the difficulties associated with preparing electrodes for transport measurement at high pressures~\cite{Mao2018Solids}.

The Hamiltonian describing the NV center ground-state is given by
\begin{align}
    \mathcal{H}&=\left[D + M_{z}\right]S_{z}^{2} + M_{x}(S_{y}^{2}-S_{x}^{2}) + M_{y} \{S_x,S_y\} &\nonumber\\
    &+ N_{x} \{S_x,S_z\} + N_{y} \{S_y, S_z\} + \gamma_{B}\bm{S}\cdot \bm{B},
    \label{eq_hamiltonian_NV}
\end{align}
where $D$ is the longitudinal zero-field splitting (ZFS) which is temperature $T$~\cite{Acosta2010Temperature, Chen2011Temperature, Doherty2014Temperature} and pressure $P$ ~\cite{Doherty2014Electronic, Steele2017Optically, Ho2020Probing, Ho2023Spectroscopic, Hilberer2023Enabling} dependent. The NV-frame quantities $M$ and $N$ can be expressed in terms of the crystal-frame stress tensor $\bm{\sigma}$ (see references~\cite{Barfuss2019Spin, Barson2017Nanomechanical, Udvarhelyi2018Spin}), and the last term is the Zeeman splitting with $\gamma_{B} = 28$~MHz/mT. The hyperfine structure is irrelevant in our measurement range.

\begin{figure}[t]
\includegraphics[width=8.6cm]{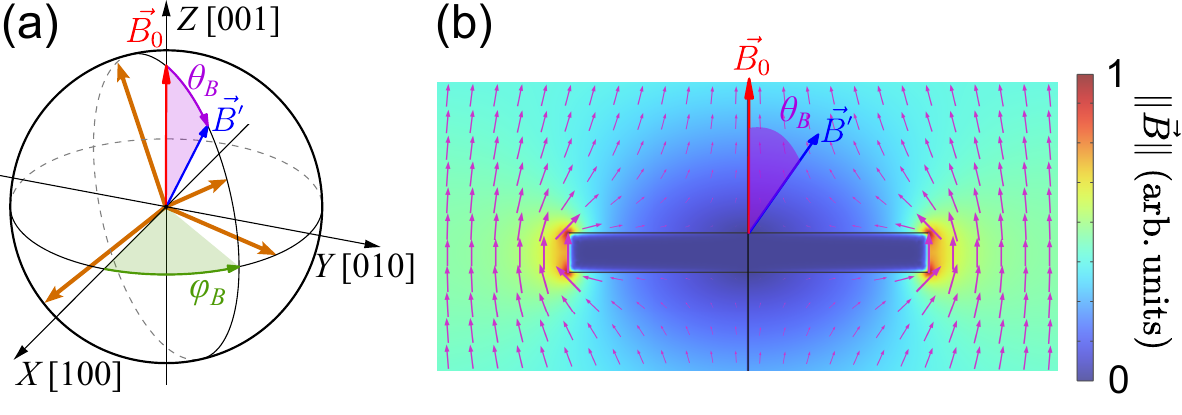}
\caption{(a) Definition of the fitting parameters with respect to the crystallographic axes of the diamond (see text). The brownish arrows represent the four NV axes. The red arrow at $T>T_{c}$ represents the applied magnetic field $\vec{B_{0}}$ along the  $[001]$, while the blue arrow $\vec{B^{'}}$ at $T<T_{c}$ illustrates the bending of the magnetic field angle (shortening of the vector is not shown). This only describes one point on the NV ensemble and is spatially dependent. (b) Illustration of the magnetic field expulsion around a superconducting cylindrical disk with a crystallographic axis $c$. The arrows show both the bending of the magnetic field vector and the decrease of its magnitude. Note that $B_{0}\parallel c\parallel[001]$.}
\label{fig:fig1}
\end{figure}

In our phenomenological model, we make a major assumption about the MW polarization applied for the NV electronic spin resonance. At temperatures above the critical temperature $T_c$, the orientation of the linearly polarized MW and the static magnetic field $B_{0}$ applied to induce the field screening align with the crystallographic $c$ axis of the sample and the [001] direction of the NV-doped diamond~\cite{Dailledouze2025Imaging}. Since the MW wavelength is much larger than the sample's dimensions, we interpret the MW magnetic field as a time-dependent amplitude modulation of the static magnetic field. Thus, the MW magnetic field is expelled in the same manner as the static magnetic field and we assume its orientation remains aligned with the expelled local static magnetic field, independent of other experimental conditions.

We then introduce eight fitting parameters to compute the resonant energies and simulate the corresponding ODMR spectrum: $\{B~\mathrm{field:}~\norm{\vec{B}}, \theta_{B}, \varphi_{B}, \epsilon_{B}$; $\mathrm{Lorentzian:}~C, \delta\nu$; $\mathrm{Pressure~ and~stress~environment:}~P, \alpha\}$. As shown in \cref{fig:fig1}(a) and \cref{fig:fig1}(b), we use spherical coordinates to represent the magnetic field with amplitude $\norm{\vec{B}}$, polar angle $\theta_{B}$, and azimuthal angle $\varphi_{B}$. The ODMR resonances are modeled by a set of Lorentzian functions with contrast $C$ and linewidth $\delta\nu$ (full width at half maximum). A complementary parameter $\varepsilon_B$ is introduced to adjust the ODMR contrast (see below). As documented in our previous work~\cite{Hilberer2023Enabling}, the stress which is applied to the NV centers deviates from a hydrostatic pressure $P$ due to the cupping effect of the anvil tip. This deviation is modeled by the $\alpha$ parameter, with $\alpha<1$.

The contrast of the associated ODMR lines for a given NV orientation in the crystal is affected by the orientation between the linear MW polarization and the NV axis, and the value $C$ of the ODMR contrast value is modulated sinusoidally with the angle between the MW magnetic field and the NV axis. In the ideal case, $C$ reaches its highest value when these directions are perpendicular and $C$ vanishes when they are parallel. To account for any possible deviation in the field misalignment, which reduces the amplitude of this modulation, we introduce the phenomenological parameter $\epsilon_{B}$ for adjusting the modulation depth (see supplemental materials).

Based on this phenomenological model, we perform constrained global fits to the experimental data by taking into account both the Hamiltonian solutions and the relative contrast of resonances. We also assume smooth evolution between adjacent positions in the magnetic field mapping and between successive temperatures. This assumption allows the fitting parameters extracted at a given position or temperature can serve as reasonable input conditions for the next fit. Therefore, we always seek a self-consistent and continuous magnetic field evolution in space and temperature.

For this work, we fixed $\alpha=0.45$ at an initial guess of $P\approx 4$~GPa. We then determine the parameter $\delta \nu$ by fitting the experimental lineshape recorded above the critical temperature $T_{c}$. Finally, the fit of the dataset is realized with six free parameters $\norm{\vec{B}}, \theta_{B}, \varphi_{B}, \epsilon_{B}, C, \mathrm{and}~P$.

The data processed here were obtained using a zero-field-cooling (ZFC) followed by field-warming (w) procedure (ZFC-w). The experiment was performed with an excitation laser power of 34.7~mW, a microwave power of 20~dBm, and an applied magnetic field $B_{0}=3~\mathrm{mT}$ along the sample $c$ axis. More experimental details are given in our previous work~\cite{Dailledouze2025Imaging} and in the supplemental materials. The widefield dataset consists of maps recorded at increasing temperatures, corresponding to about three million data points. This dataset was reduced to about twenty-five thousand data points by selecting one vertical line cut, one horizontal line cut, and a square around the superconducting sample.

\begin{figure}[t]
\includegraphics[width=8.6cm]{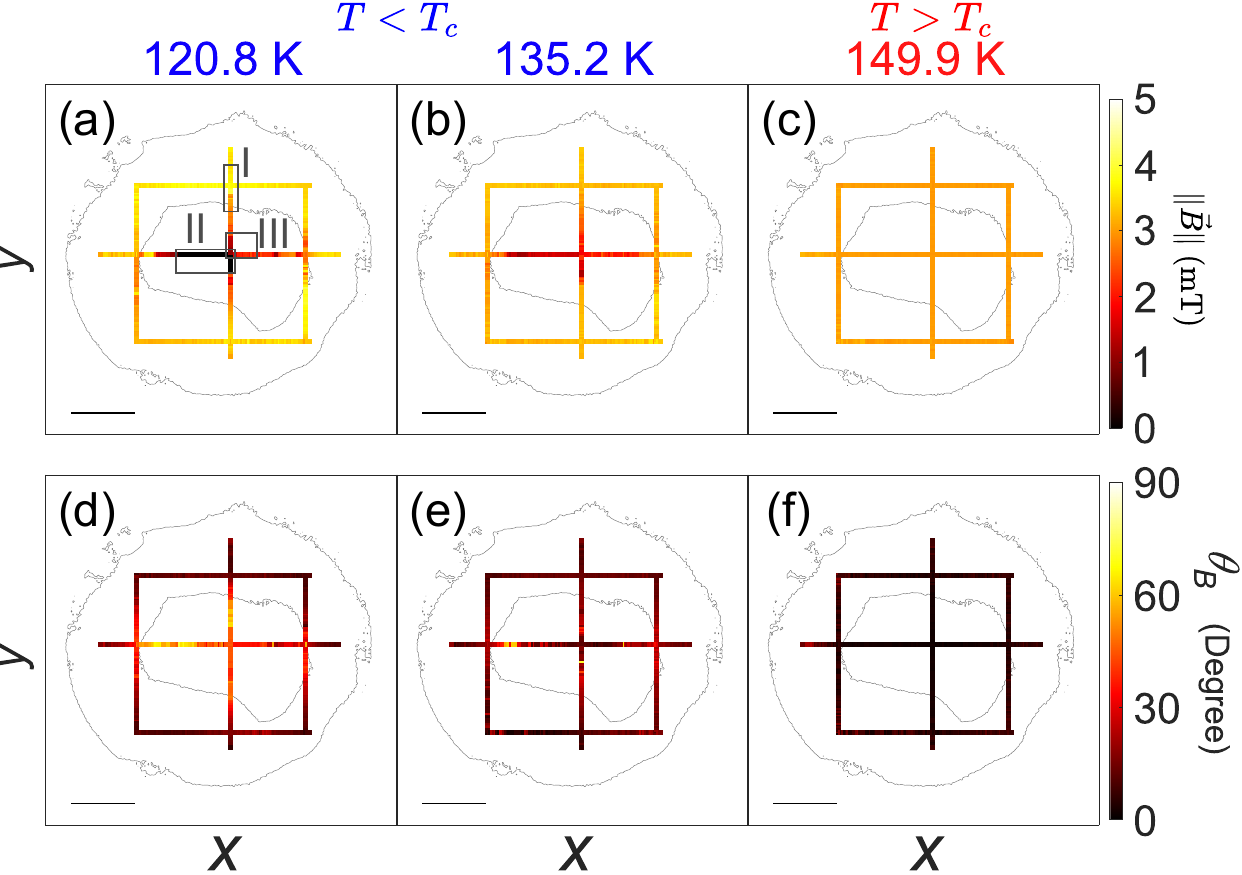}
\caption{(a-c) Maps of the magnetic field magnitude $\norm{\vec{B}}$ and (d-f) maps of the polar angle $\theta_{B}$ obtained at temperatures of 120.8~K, 135.2~K, and 149.9~K, respectively. The field screening results in a higher magnetic field magnitude surrounding the sample, and a lower magnetic field and a larger polar angle towards the center of the superconducting Hg-1223 microcrystal. The inner and outer contours outline the sample geometry and the pressure chamber, respectively. The scale bar is 25~$\mu$m. Regions I, II, and III are labeled for three key observations (see text).}
\label{fig:fig2}
\end{figure}

In \cref{fig:fig2}(a-c) and \cref{fig:fig2}(d-f), we plot the magnitude $\norm{\vec{B}}$ of the magnetic field and its polar angle $\theta_{B}$, as determined by the fitting procedure applied to the dataset recorded at temperatures of 120.8~K, 135.2~K, and 149.9~K, respectively. The critical temperature $T_{c}$ is approximately 138~K; At $T<T_{c}$, $\norm{\vec{B}}$ and $\theta_{B}$ exhibit a strong spatial dependence due to the field screening, whereas at $T>T_{c}$, these parameters become nearly uniform. Additionally, our quantitative analysis yields three key observations. First, in region I, the sharpest change in the magnetic field profile at the top part of the vertical line cut correlates to the sharpest edge of the superconductor (see supplemental materials). The irregular shape of the sample causes the difference in the field profile observed at both ends. Second, region II has the lowest magnetic field. This is presumably due to the shortest distance between the NV centers and the sample surface, resulting in the strongest magnetic field reduction. Third, region III (inside the sample) exhibits finite residual magnetic fields at $T<T_{c}$, which is consistent with the defect in the sample identified in the analysis done in our previous work~\cite{Dailledouze2025Imaging}. Magnetic flux tends to penetrate through any defect in a superconductor, thus giving a finite residual magnetic field at that location.

\begin{figure*}[t]
\includegraphics[width=\textwidth]{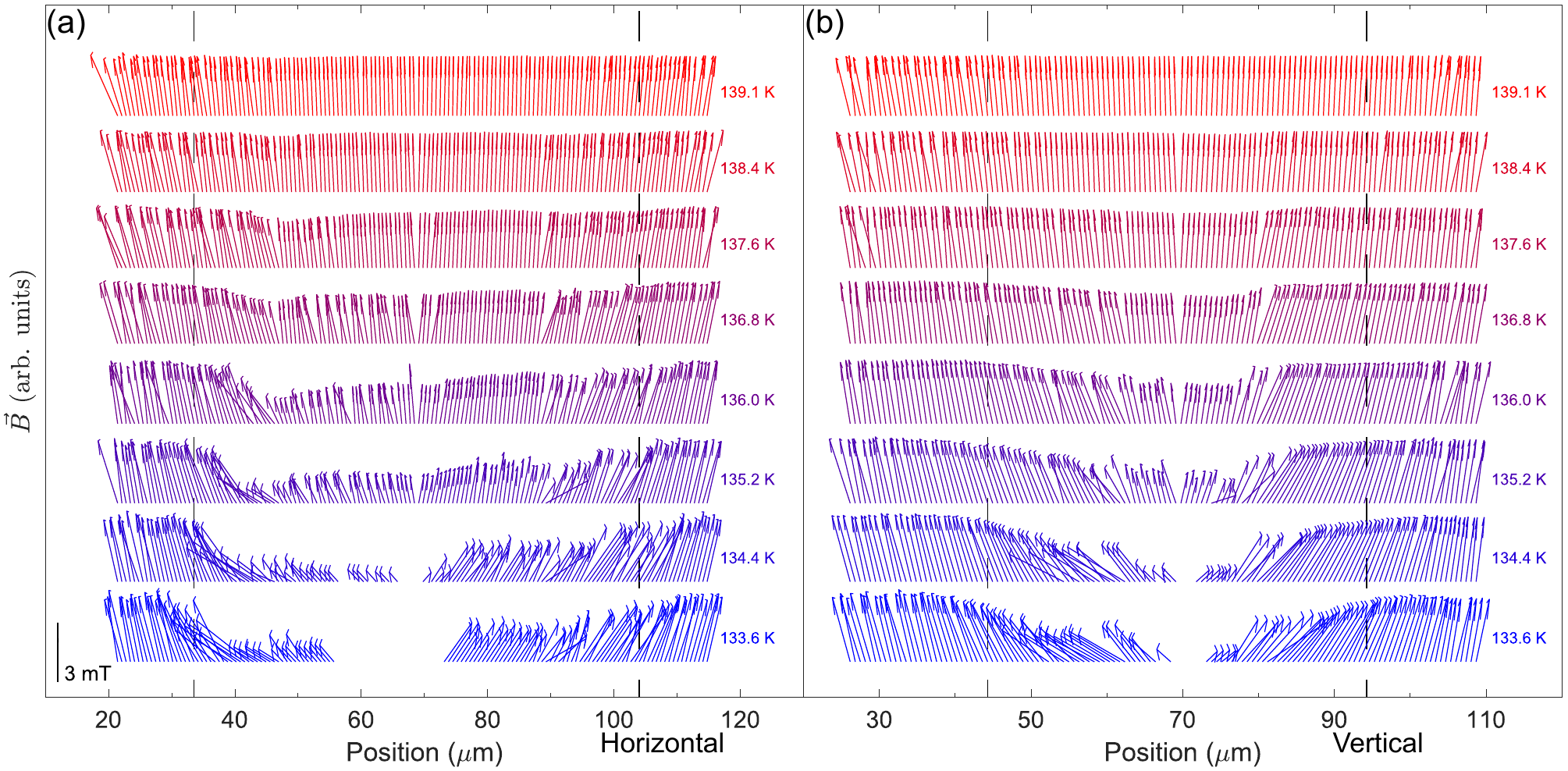}
\caption{Magnetic field reconstruction of the field screening at different temperatures for (a) horizontal and (b) vertical line cuts. We focus on the results near the critical temperature $\sim 138$~K, otherwise, the data show negligible variation. Each arrow represents the magnitude $\norm{\vec{B}}$ and the polar angle $\theta_{B}$ of the magnetic field. Due to the strong expulsion ($B\approx0$~mT) of the magnetic field around the center of the Hg-1223 sample in its superconducting state, no magnetic field vectors could be drawn at low temperatures in the corresponding parts of the graphs (also see supplemental materials). When approaching the $T_{c}$, from 135.2~K to 137.6~K, the change in the magnetic field is rapid. A reference magnetic field vector with an amplitude 3~mT along the $c$-axis of the sample is drawn. The dashed black lines on the background roughly outline the width of the sample.}
\label{fig:fig3}
\end{figure*}

By exploiting this detailed description of the magnetic field magnitude $\norm{\vec{B}}$ and polar angle $\theta_{B}$, we can reconstruct the expulsion of the magnetic field vector from the superconductor as an alternative representation of \cref{fig:fig2}. Here, we utilize a vector plot to directly visualize the field screening. The resulting reconstruction of the horizontal and vertical linecuts are displayed in \cref{fig:fig3}(a) and \cref{fig:fig3}(b), respectively. The magnetic field vectors vary with the distance from the center of the superconductor, and also from the Meissner state to the normal state. At low temperatures, the reduction of the magnetic field and a large turn in the angle indicate a strong field screening. As the temperature increases, the vectors near the center gradually point back to the sample $c$ axis with increasing magnitudes, while those near the edge gradually decrease in magnitude with a small change in angle. Both features provide an indisputable proof of the diamagnetism associated with the Meissner effect. Furthermore, the changes in the magnetic field is rapid when approaching $T_{c}$, commonly understand as vortex dynamics in the phrase transition. We notice that this can also be understood by Brandt's model (see below).


\begin{figure}[t]
\includegraphics[width=8.6cm]{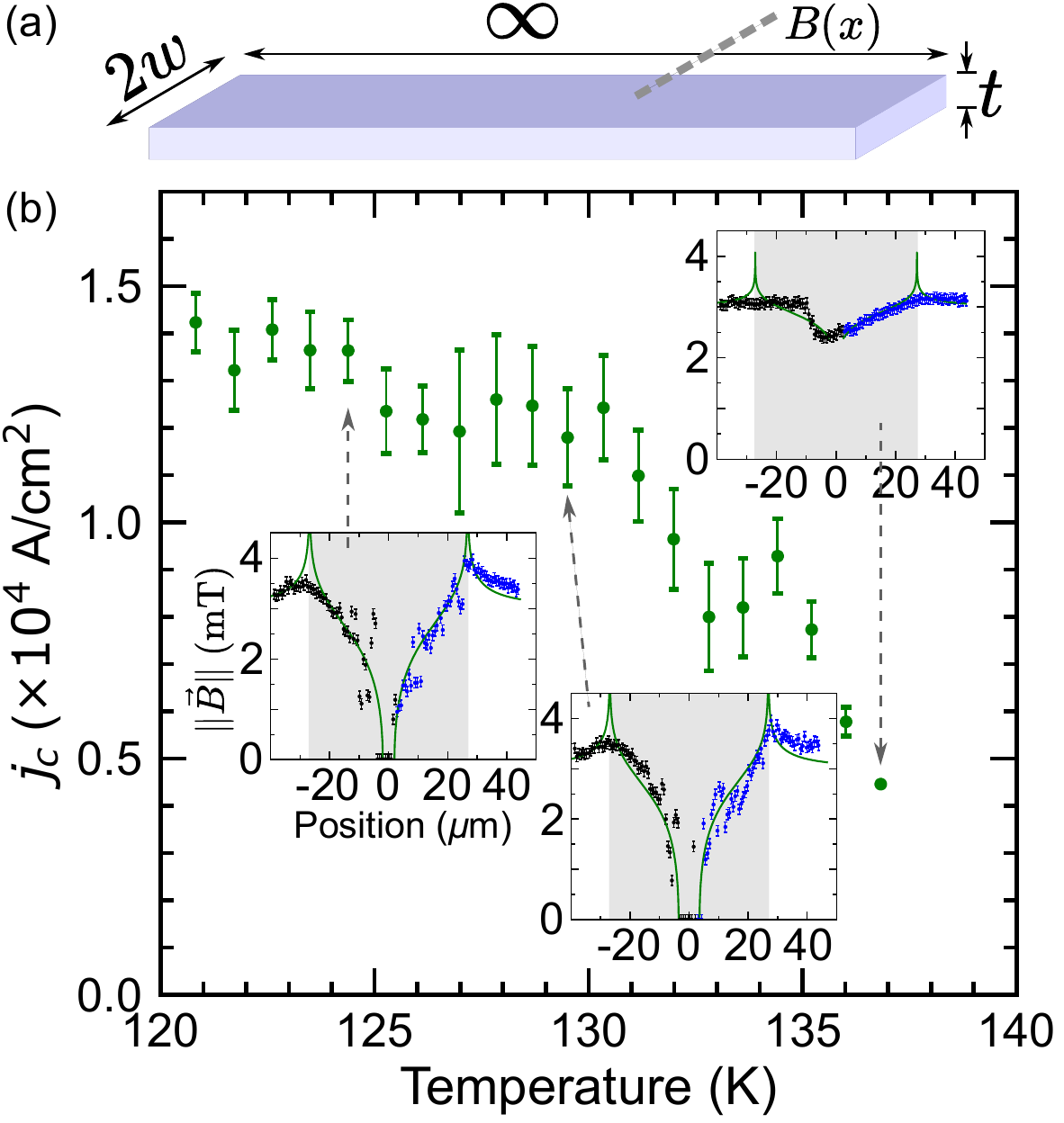}
\caption{(a) Model of a semi-infinite superconducting slab, with a width $2w$ in the transverse direction $x$ and a thickness $t$. Brandt's model is used to calculate the magnetic field profile along the dashed-gray line cut $B(x)$ in order to determine the critical current density $j_{c}$. (b) Temperature-dependent critical current density $j_{c}$ of Hg-1223 obtained using Brandt's model. The critical current density value is approximately $1.5\times10^4\, {\rm A} \cdot {\rm cm} ^{-2}$ at 120~K and 4~GPa. Inset: Fits of the magnitude of the magnetic field along the transverse $x$ direction, for three temperatures below $T_c$: 124.4~K, 129.5~K, and 136.8~K. The sharpest edge on the positive $x$-side (blue markers) is used to obtain the best fit, shown by the green line. The corresponding values obtained from this best fit are then plotted for the negative $x$-side (black markers). Due to the finite spatial resolution and the absence of sharp sample edges, changes in the magnetic field on the edges are smoothed out. The dashed black lines on the background indicate the width of the sample. Note that we do not account for the standoff distance between NV layer and Hg-1223, which has a minimum distance of 20~nm due to the NV implantation depth.}
\label{fig:fig4}
\end{figure}

In addition to reconstructing the magnetic field profile around the superconducting sample, we use Brandt's model~\cite{Brandt1993TypeII} to calculate the critical current density $j_{c}$, which is the maximum current density that a type-II superconductor can withstand before losing its superconducting properties. As shown in \cref{fig:fig4}(a), the simplified geometry used in this model corresponds to a semi-infinite superconducting slab with a width $2w$ and a thickness $t$, where $w\gg t$. Brandt's model then yields the magnetic field profile $B(x)$ for an external magnetic field $B_{0}$ applied perpendicular to the slab. This creates superconducting currents associated with a constant critical current density $j_{c}$ so that :
\begin{equation}
B(x) = 
\begin{cases}
0, &\mathrel{\hphantom{\beta<}}\abs*{x}<\beta,\\
\frac{\mu_{0}j_{c}t}{\pi}\mathrm{arctanh}\left( \frac{\sqrt{x^{2}-\beta^{2}}}{\gamma\abs*{x}} \right), &\beta<\abs*{x}<w,\\
\frac{\mu_{0}j_{c}t}{\pi}\mathrm{arctanh}\left( \frac{\gamma\abs*{x}}{\sqrt{x^{2}-\beta^{2}}} \right), &\mathrel{\hphantom{\beta<}}\abs*{x}>w,
\end{cases}
\label{eq:Brandt}
\end{equation}
where $\mu_{0}$ is the vacuum permeability and
\begin{align}
\beta &= \frac{w}{\mathrm{cosh}\left[\pi B_{0}/(\mu_{0} j_{c} t)\right]},\\
\gamma &= \mathrm{tanh}\left[\pi B_{0}/(\mu_{0} j_{c} t)\right].
\end{align}
The parameter $\beta$ can be interpreted as the lateral penetration depth of the magnetic field in the superconductor and the parameter $\gamma$ gives a ratio between the external magnetic field $B_{0}$ and the surface magnetic field induced by the critical sheet current $J_{c}=j_{c}t$. Note that the critical current density $j_{c}$ is the only fitting parameter in these equations.

The Brandt's model has previously been applied to datasets obtained by NV magnetometry without knowledge of the detailed magnetic field profile~\cite{Paone2021All, Ho2025Studying}. It was then shown that during the phase transition, the NV splitting alone is insufficient for a complete description~\cite{Ho2025Studying}. Here, we take a step forward by using Brandt's model to fit the reconstructed magnetic field profile even though the sample geometry does not perfectly match that of an ideal slab as considered in the model. The temperature-dependent critical current density $j_{c}$ determined using this method is summarized in \cref{fig:fig4}(b) (also see supplemental materials). The insets show the magnetic field profile of the line cut (blue and black markers), together with Brandt's model fit (green line) at different temperatures. Although the experimental data are slightly scattered, they largely follow Brandt's model. Since the magnetic field resembles that produced by a current sheet, the sharpest change in the magnetic field is observed at the edges, where the magnetic field magnitude is highest and decreases rapidly inside the superconductor. As expected, we also observe an increase in the critical current density $j_{c}$ when cooling deeper into the superconducting state. Another important observation is that the parameter $\beta$ exhibits an exponential sensitivity in the low-$j_{c}$ regime. Consequently, near $T_{c}$ where $j_{c}$ is small, a modest change in the temperature lead to a significant redistribution of the magnetic field around the superconductor. Whereas for $T\ll T_{c}$ where $j_{c}$ is large, $\beta$ approaches a saturated value, rendering the magnetic field screening only weakly dependent on temperature over a wide low-temperature range. For the Hg-1223 sample at approximately 4~GPa, the value of the critical current density is $j_{c}\approx1.5\times10^4$~A$\cdot$cm$^{-2}$ at 120~K, in excellent agreement with literature values of $\sim1\times10^{4}-10^{5}$~A$\cdot$cm$^{-2}$ at slightly lower temperatures and ambient pressure~\cite{Tsabba1996Giant, Schilling1993Magnetization, Gapud1997Effects, Yun1996Superconductivity, Yun1996Growth}.

We found a finite pressure variation in our dataset (see supplemental materials). In particular, a direct contact between the sample and the diamond creates a higher local pressure. Therefore, the local pressure environment at even higher pressures should be carefully considered. Pressure mapping also underscores the power of widefield measurements in high-pressure studies.


The Gaussian profile of the excitation laser and the non-uniform MW magnetic field induce some spatial dependence of the data, leading to errors in the reconstructed magnetic field. We estimate that the error in the magnetic field magnitude $\norm{\vec{B}}$ is below $\pm 0.1$~mT whereas the error in the polar angle $\theta_{B}$ is approximately $\pm 5^{\circ}$. However, the azimuthal angle $\varphi_{B}$ depends heavily on the polar angle $\theta_{B}$ and could be easily varied within $\pm10^{\circ}$. Using machine learning tools and/or artificial intelligence methods to efficiently analyze the data should lead to a more refined model and eventually reconstruct the magnetic field vector everywhere.

In summary, to investigate the diamagnetism associated with the Meissner effect of a superconducting Hg-1223 microcrystal under pressure, we fitted the resonance lineshapes in widefield ODMR spectra by parameters of the corresponding Hamiltonian and the contrasts with a phenomenological model. This model, which assumes that the expelled static magnetic field and the MW excitation magnetic field remain aligned, allowed us to reconstruct the magnetic field distribution. We showcase how the magnetic field vector bends spatially around the superconductor, and also how this distortion evolves with temperature. We finally apply Brandt's model to determine the temperature-dependent critical current density $j_{c}$, the result being in good agreement with previous measurements at ambient pressure. With the ability to study lower and upper critical fields $H_{c1}$ and $H_{c2}$ under pressure~\cite{Yip2019Measuring}, we conclude that by solely using NV magnetometry, one can determine these three critical parameters for a superconductor under pressure.

See the supplementary material for details on the experimental details, phenomenological model fitting, pressure variation, and extended data on the linecut and binned spatial mapping.

We acknowledge Paul Loubeyre for fruitful discussions and comments. We thank Anne Forget and Dorothée Colson for providing us with the Hg-1223 sample. This work has been funded by the Region \^Ile-de-France in the framework of the DIM QuanTIP, by the ANR with the ESR/EquipEx+ program (Grant No. ANR-21-ESRE-0031) and by the European Research Council with the ERC Advanced Grant ``QPRESSE'' (No. 101142682). J.-F. Roch acknowledges support from Institut Universitaire de France.

\section*{Author Declarations}
\subsection*{Conflict of Interest}
The authors have no conflicts to disclose.

\subsection*{Author Contributions}
\textbf{Kin On Ho:} Conceptualization (equal); Data curation (equal); Formal analysis (equal); Methodology (supporting); Writing – original draft (lead); Writing – review \& editing (lead). \textbf{Cassandra Dailledouze:} Data curation (equal); Formal analysis (equal); Methodology (supporting); Writing – review \& editing (supporting). \textbf{Martin Schmidt:} Conceptualization (equal); Formal analysis (supporting); Methodology (lead); Supervision (lead); Writing – review \& editing (supporting). \textbf{Loïc Toraille:} Formal analysis (supporting); Writing – review \& editing (supporting). \textbf{Marie-Pierre Adam:} Resources (equal); Writing – review \& editing (supporting). \textbf{Jean-François Roch:} Resources (equal); Supervision (supporting); Writing – review \& editing (supporting).
\section*{Data Availability}
The data that support the findings of this study are available from the corresponding author upon reasonable request. 

\bibliography{references}

\end{document}